\documentclass[12pt]{article}
\usepackage{graphicx}
\usepackage{psfrag}
\usepackage{amsmath, amssymb}
\usepackage{amsthm}
\usepackage{endnotes}
\usepackage{natbib}

\usepackage{mathpazo} 
\usepackage[scaled]{helvet} 
\usepackage{courier} 
\normalfont

\newcommand{\field}[1]{\mathbb{#1}}

\newcommand{\R}{\field{R}}

\newcommand{\mc}{\mathcal}

\newcommand{\eod}{\vspace{0in}\hspace{\fill}{$\diamondsuit$}}
\newcommand{\eop}{\vspace{0in}\hspace{\fill}{$\Box$}}
\newcommand{\be}{\begin{equation}}
\newcommand{\ee}{\end{equation}}
\newcommand{\bes}{\begin{equation*}}
\newcommand{\ees}{\end{equation*}}

\theoremstyle{definition}

\newtheorem{defn}{Definition}

\newtheorem{proposition}{Proposition}
\newtheorem{theorem}{Theorem}

\theoremstyle{remark}

\newtheorem*{rems}{Remarks}

\title{Mathematical analysis of Soros's\\theory of reflexivity}
\author{C.P. Kwong\thanks{The Chinese University of Hong Kong, Hong Kong.\newline
\emph{Address for correspondence}: Department of Mechanical and Automation Engineering, The Chinese University of Hong Kong, Shatin, N.T., Hong Kong; email: cpkwong@mae.cuhk.edu.hk}}
\date{November 27, 2008}      

\begin{document}
\maketitle
\abstract{The mathematical model proposed by George Soros for his theory of reflexivity is analyzed under the framework of discrete dynamical systems.  We show the importance of the notion of fixed points for explaining the behavior of a reflexive system governed by its cognitive and manipulative functions.  The interrelationship between these two functions induces fixed points with different characteristics, which in turn generate various system behaviors including the so-called ``boom then bust" phenomenon in Soros's theory.


}

\section{Introduction}  The theory of reflexivity is a theoretical construct of George Soros's philosophy on prices in financial markets.   The theory was first documented by Soros in his \textit{The Alchemy of Finance}~\citep{sor87}, expanded in \textit{Open Society}~\citep{sor00}, and
reiterated, using the recent credit crisis as example, in his latest work \textit{The New Paradigm for Financial Markets}~\citep{sor08}.  From the outset Soros does not believe, that the methods of natural sciences can be applied to the study of social sciences such as economics.  Simply put, in his own words, there is no hope to obtain Newton's laws for finance or economics.  Specifically he defies price theories grounded on equilibrium as can be noticed by the title of the first section of the chapter ``The Theory of Reflexivity" in his \textit{The Alchemy of Finance}, namely ``Anti-equilibrium". His basic argument is that human beings will inevitably participate in any pricing process.  Since their perspectives on price, and hence their reactive actions which influence the future price, are ever changing, there is no room for equilibrium but fluctuations.    Exactly from this premise Soros proposed his theory of reflexivity.  Though his approach to reflexivity has been mainly philosophical, Soros did attempt to model his theory {\bf\emph{mathematically}} using the following pair of equations from the very beginning~\citep{sor87}:
\be\label{reflex_eqns}\begin{cases}y=f(x),\\x=\phi(y).\end{cases}
\ee
He calls $f$ the {\bf cognitive function} in which ``the participants' perceptions depend on the situation" and $\phi$ the {\bf manipulative function} in which ``the situation is influenced by the participants' perceptions".\endnote{Originally Soros called $\phi$ the {\bf participating function}.  He recently considered ``manipulative function" a better substitute~\citep[p.3]{sor08}.  We share his view.}  The following quote from Soros~\citep[p.17]{sor08} may give some hints on his attitude towards the use of mathematics in economics:\begin{quote} ``\textit{I was not very good at math, and that led me to question the assumptions on which the mathematical models of economists were based."}\end{quote}

According to the inventor himself the theory of reflexivity has received no serious attention from the academia since its inception---it is ``totally ignored in departments of economics"~\citep[p.20]{sor08}.  Soros attributes this phenomenon to the imprecision in his formulations, and ``As a result, the professionals whose positions I challenged could dismiss or ignore my arguments on technical grounds without giving them any real consideration"~\citep[p.20]{sor08}.   We cannot agree him more except that Soros has indeed formulated his theory by a {\bf\emph{precise}} mathematical model, i.e., Equation~\eqref{reflex_eqns}.  The real problem arises from the meagre attention of this equation by Soros himself or others.  For instance all Soros's published writings are narrative in nature and neither other representative reviews on the theory of reflexivity have addressed this fundamental equation~\citep{cro&str97,bry02}.  Soros may be excused from the duty of performing the mathematical analysis because he ``was not very good at math".  But for other mathematically inclined researchers in finance or economics, the lack of such an interest is somewhat puzzling since Equation~\eqref{reflex_eqns} is very amenable to mathematical analysis as we will show in this paper.  The article, published by Birshtein and Borsevici~\citep{bir&bor02}, is perhaps the only exception.  There the authors make explicit the recursive nature of Equation~\eqref{reflex_eqns} and then use graphical analysis to show that equilibria do exist in the reflexive model, a result in contrast to Soros's dubiousness on equilibrium theory.  Nevertheless, their analysis is incomplete and the present paper can be considered as the continuation of their efforts.  We believe that such an analysis is indispensable to a real understanding of the theory of reflexivity as formulated by Soros, no matter the results are supporting his claims, which go beyond finance to reach other social science subjects, or otherwise. This is exactly the purpose of our study.

At this junction it may be worthwhile to consider the role of mathematical modeling in studying economics in order to justify our present effort.  The topic is big and controversial~\citep{mir86,deb91,bee&kan91,vel05}.  However, our view is simple (and of course, nonsingular) enough to be expressed within a paragraph.  We admit that what social scientists are trying to explore are systems of extreme complexity.   Not only that the number of variables can be huge but also complicated human behaviors may be involved.  However, it should not, in our opinion, to preclude the possibility of using some ``good" mathematics to effect modeling and analysis of these complex social systems as we do for physical or biological systems.  The mathematics so used need not be very advanced and abstract to be effective.  The works of Nobel Laureates Daniel Kahneman~\citep{kah&tve79} and Thomas C. Schelling~\citep{sch06} for modeling human behaviors are two typical examples.  It is important to learn from these examples, that in many cases the effectiveness of mathematics in an applied science lies not in providing precise computations of variables but sufficiently accurate {\bf\emph{qualitative}} descriptions of phenomena.  This is particularly true when we study systems that cannot be modeled exactly.

We divide our paper into several sections.  In Section 2 we set up a dynamical system model for the study of Equation~\eqref{reflex_eqns} and use a simple example to show the basic characteristics of this model.  In Section 3 we analyze the model behavior in detail.  Although the mathematical results we shall use are well documented, we will nevertheless supply their full descriptions for readers who are not familiar with these topics.  Finally we conclude our work in Section 4 together with some additional comments.

\section{The System Model for Reflexivity}\label{section_model} Equation~\eqref{reflex_eqns} can be depicted by the following block diagram which shows clearly the causal relationships between the variables $x$ and $y$:

\vspace{0.5cm}
\begin{center}
\begin{figure}[!h]\centering
\psfrag{x}[][]{$x$}\psfrag{y}[][]{$y$}
\psfrag{f}[][]{$f$}\psfrag{p}[][]{$\phi$}
\includegraphics{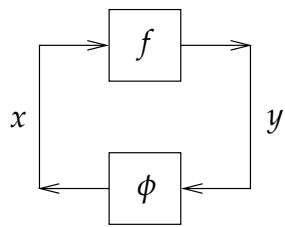}
\caption{The system-theoretic model of reflexivity.}\label{fig_reflex_eqns}
\end{figure}\end{center}
\vspace{0.5cm}

\noindent The block diagram also suggests the use of ``systems theory" to study reflexivity.\endnote{Very roughly speaking systems theory is a theoretical framework for studying how an object interacts with its environment through interchanging energy and information.  It has been applied to many fields of study including biology, engineering, and social sciences.}  For instance, in the language of systems theory, $x$ is the {\bf input signal} to the cognitive function $f$ and $y$ is the {\bf output signal} of the function (treated as a system).  The diagram represents a {\bf feedback system} since the output $y$ of $f$ is fed back to its input after $y$ is mapped by the manipulative function $\phi$.  We can also say that the output $x$ of $\phi$ is fed back to its input after $x$ is mapped by the cognitive function $f$.  Because of this symmetry $f$ takes the same role of $\phi$ when it is considered as a device which modifies the feedback signals.  For easy reference in the future we use $\mc{S}$ to label this system.

Given a system $\mc{S}$ of Figure~\ref{fig_reflex_eqns} with specific cognitive and manipulative functions, the values of $x$ and $y$ are not arbitrary but governed by the feedback connection of these two functions.  Indeed they are simply the solution(s) of Equation~\eqref{fig_reflex_eqns}.  The significance of this constraint on variable values, which is introduced by the feedback configuration or equivalently Equation~\eqref{fig_reflex_eqns}, is illustrated by the following simple yet elucidative example.

Let $x$ be the exchange rate of a particular currency (say Euro versus USD) and $y$ be its domestic interest rate.  According to the interpretation of Soros, the cognition function $f$ would take the role of understanding the current exchange rate $x$ and then would form an opinion on the interest rate $y$.  The manipulative function $\phi$ reacts to this value of $y$ and outputs a new value of exchange rate.  For the sake of further analysis of the reflexive system $\mc{S}$ we assume the existence of these two functions without detailing how these functions can be derived.\endnote{There are well-known theories on the relationship between exchange rates and interest rates.  For instance, the ``Chicago" theory~\citep{bil78} proposes ``a positive relationship between the exchange rate and the nominal interest rate differential" while the ``Keynesian" theory~\citep{dor76} suggests a ``negative relationship between the exchange rate and the nominal interest rate differential"~\citep{fra79}.} The process continues as dictated by the feedback loop of Figure~\ref{fig_reflex_eqns}.

From the above example it is easy to see that the values of $x$ and $y$ evolve with time step-by-step starting from some initial value of $x$.  Suppose we label this initial value as $x_0$, then we obtain subsequently two sequences of numbers $x_1,x_2,\dotsc$ and $y_1,y_2,\dotsc$.  Clearly the subscripts $1,2,\dotsc$ are used to index time.  Moreover, we have $y_0=f(x_0),x_1=\phi(y_0),y_1=f(x_1),x_2=\phi(y_1)$, and so on.  Thus we can write two general expressions for the respective evolutions of  $x$ and $y$ (with time) as
\be\label{recur_f} y_i=f(x_i),\quad i=0,1,2,\dotsc \ee
and
\be\label{recur_phi} x_{i+1}=\phi(y_i),\quad i=0,1,2,\dotsc. \ee
In this way we turn $\mc{S}$ into a {\bf discrete dynamical system} by ``iteration of functions".  Following the tradition of dynamical systems theory we call the pair $(x_i,y_i)$, a point on the Euclidean plane, the {\bf state} of $\mc{S}$ and the set of points $\{(x_0,y_0),(x_1,y_1),(x_2,y_2),\dotsc\}$ the (forward) {\bf orbit} of $(x_0,y_0)$ under $\mc{S}$.  Dynamical systems of this kind have been extensively studied (see, for example, \citep{hal&koc96}, \citep{hol96}, and \citep{rob04}) and this is why we claim in the Introduction that Equation~\eqref{reflex_eqns} is very amenable to mathematical analysis.  In the next section we will frequently apply existing results from discrete dynamical systems to our study.  However, before moving on to the formal analysis of the dynamics of $\mc{S}$, we would like to use the exchange rate example to motivate our research along this direction.

As above we let $x_0$ be the initial exchange rate of a currency.  The cognition function $f$ takes $x_0$ as input and outputs the interest rate $y_0=f(x_0)$.  Suppose the manipulative function $\phi$ reacts to $y_0$ in a way such that $x_1=\phi(y_0)=x_0$, then we see immediately that the value of $x_i$ as well as the value of $y_i$ will remain {\bf\emph{unchanged}} for all subsequent $i$, i.e., $x_i=x_0, y_i=y_0, i=1,2,\dotsc$.  Mathematically we call the pair $(x_0,y_0)$ the "fixed point" of the dynamical system $\mc{S}$, which we shall formally define and study in the next section.  The concept of fixed point is significant in applying the theory of reflexivity.  This is because the existence of a fixed point (or fixed points) implies that both parties who determine the respective cognitive and manipulative functions do {\bf\emph{agree}} with the one-one correspondence between the elements of this particular pair (or pairs) of $x$ and $y$.  For the exchange rate example that will mean an agreement in which a particular exchange rate should correspond to a particular interest rate.  The system thus reaches an equilibrium.  In the extreme case the parties agree on all pairs of $x$ and $y$.  This will happen when the manipulative function $\phi$ is the ``inverse" of the cognitive function $f$, or vice versa.  It is worth to note, before giving the details, that even the initial state does not begin at a fixed point but in its vicinity, the subsequent state may approach and settle at the fixed point after several iterations.  In real terms the parties have different opinions in the beginning but eventually agree with each other.  Alternatively the state leaves the fixed point and does not return.  The dynamical behavior of $\mc{S}$ as influenced by its fixed points is our immediate topic of study.

\section{The System Behavior}
\subsection{Background Mathematics}
We begin with the definition of a function, assuming that the readers already know what is a set.  We must stress again that all the mathematics presented here are standard (see, for example,~\cite{gas&nar98}), and their inclusion here is for easy reference only.
\defn Let $A$ and $B$ be sets.  A {\bf  relation} $R$ on $A$ and $B$ is a
collection of ordered pairs $(a,b)$ such that $a\in A$ and $b\in B$.\eod

\defn Let $A$ and $B$ be sets.  A {\bf  function} $f$ from $A$ to $B$ is a relation
$R$ on $A$ and $B$ such that for every $a\in A$ there is one and only
one pair $(a,b)\in R$.  We call $A$ the {\bf  domain} of the function
and $B$ its {\bf  codomain}.  A function is also called a {\bf map}.\eod

\medskip
Quite often we write $f\colon A\rightarrow B$ or
$A\xrightarrow{f} B$ for ``a function $f$ from $A$ to $B$," and write
$a\mapsto f(a)$ to indicate that "$a$ is mapped to $f(a)$ under $f$".
\defn Let $f\colon A\rightarrow B$ be a function.  The {\bf  range} of $f$ is the set
\bes
\{b\in B\,\big|\,\text{there exists} \ a\in A \ \text{such that} \ f(a)=b\}.
\ees

\vspace{-0.4cm}
\eod

\medskip
\noindent Note that the range of $f$ is not necessarily the whole codomain $B$.
For example, the function represented by
$y=\sin(x)$ has the set of real numbers $\R$ as both its domain and codomain.  However, the range of this function is only the closed interval $[-1,1]$.
Another example is the function $f$ represented by $y=\log(x)$, of which the domain is the set
$\{x\in\R\,\big|\,x>0\}$.  If we denote this set by $\R^{+}$, we may write
$f\colon \R^{+}\rightarrow\R$.

\defn A function $f\colon A\rightarrow B$ is {\bf  injective} if whenever
$(a_{1},b)\in f$ and $(a_{2},b)\in f$, then $a_{1}=a_{2}$.  It is
called {\bf  surjective}, or {\bf  onto} if whenever $b\in
B$, then there exists an $a\in A$ such that $(a,b)\in f$.  Finally,
the function is called {\bf  bijective}, or {\bf  one-one} if it
is both injective and surjective.\eod

\medskip
\noindent For instance the function represented by $y=\sin(x)$ is not injective and the function
represented by $y=\log(x)$ is bijective (one-one).  As a consequence,
the ``inverse" of this latter function exists, which is the anti-log.

\defn Let $f\colon A\rightarrow B$ be a bijective function.  Then the {\bf  inverse
function} $f^{-1}$ of $f$ is the function $f^{-1}\colon B\rightarrow A$ defined by
$f(a)\mapsto a$ for every $f(a)\in B$.\eod

\medskip
Let $A,B,$ and $C$ be sets and $f,g$ be functions $A\xrightarrow{f} B$ and
$B\xrightarrow{g} C$.  We may then write
\bes
A\xrightarrow{f} B\xrightarrow{g} C
\ees
and introduce a function $A\xrightarrow{h} C$.  The function $h$ is called the
{\bf  composition} of the functions $f$ and $g$.  In this case $h$ is called a {\bf composite function}.  We write $h=g\circ
f$.  For every $a\in A$, we have
\bes a\mapsto h(a)=(g\circ
f)(a)=g(f(a)).
\ees Note that the order of composition is
important---$f\circ g$ may not even make sense (for example when the domain of $f$ does not include the range of $g$) and if it exists, is in general {\bf\emph{not}} identical to $g\circ f$.

\medskip
Differential calculus plays a crucial role in the study of iteration of functions.
\defn Let $f\colon \R\rightarrow\R$ be a function with domain $D$ and let $a\in\R$.  We say that $f$ has a {\bf limit $A$ as $x$ tends to} $a$, written as
\bes \lim_{x\to a}f(x)=A,
\ees
if, for $\epsilon >0$, there is a $\delta >0$ such that for every $x\in D$,
\bes 0<|x-a|<\delta\quad\text{implies}\quad |f(x)-A|<\epsilon.
\ees

\vspace{-0.4cm}
\eod
\defn Let $f\colon \R\rightarrow\R$ be a function with domain $D$.  We say that $f$ is {\bf continuous at a point} $a\in D$ if, for every $\epsilon >0$, there is a $\delta >0$ such that
\bes 0<|x-a|<\delta\quad\text{implies}\quad |f(x)-f(a)|<\epsilon.
\ees
A function is said to be {\bf continuous} if it is continuous at every point in $D$.\eod
\defn Let $f\colon \R\rightarrow\R$ be a function with domain $D$ and $a\in\R$ be a point in the interval $(a-c,a+c)\subseteq D$ for some $c>0$.  We say that $f$ is {\bf differentiable at} $a$ if the following limit \bes \lim_{x\to a}\frac{f(x)-f(a)}{x-a},
\ees exists.  The limit, if exists, is called the {\bf derivative of $f$ at} $a$ and is labeled $f'(a)$.  A function is said to be {\bf differentiable} if it is differentiable at every point in $D$.\eod

\medskip\noindent Geometrically $f'(a)$ is the slope of the tangent line to $f$ at $a$, which can be positive or negative.  The absolute value of this slope is written as $|f'(a)|$.

\medskip
\subsection{Fixed Points}
We are now ready to go into the details of the dynamic behavior of the system $\mc{S}$.  First, the idea of a ``fixed-point", which we sketch in Section 2, is formally defined.

\defn The pair of real number $(\bar{x},\bar{y})$ is called a {\bf fixed point of} $\mc{S}$ if
\bes f(\bar{x})=\bar{y}\quad\text{and}\quad\phi(\bar{y})=\bar{x}.\ees

\vspace{-0.4cm}
\eod

\begin{center}
\begin{figure}[!h]\centering
\psfrag{x}[][]{$x$}\psfrag{y}[][]{$y$}
\psfrag{xb}[lc][]{$\bar{x}$}\psfrag{yb}[][]{$\bar{y}$}
\psfrag{f(x)}[][]{$f(x)$}\psfrag{p(y)}[][]{$\phi(y)$}
\includegraphics[scale=1]{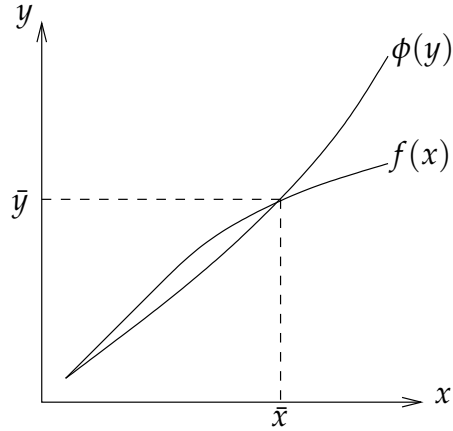}
\caption{The idea of a fixed point.}\label{fig_graph_2_fns}
\end{figure}
\end{center}

\bigskip The graph of a function usually gives a lot of information and it is even more true when we investigate the dynamics of $\mc{S}$ that consists of two functions $f$ and $\phi$ configured in the present way.  In particular, if we plot $f$ and $\phi$ on the same $x$-$y$ plane as in~Figure~\ref{fig_graph_2_fns}, {\bf\emph{the intersections of $f$ and $\phi$, if exist, are exactly the fixed points.  If the functions are inverses of each other, then their graphs overlap at all their defining points.}} Furthermore, the dynamics of $\mc{S}$, i.e., the evolution of $x$ and $y$ with time, can be easily visualized by using this graphical representation as we will show in Subsection \ref{subsection_behavior}.

\vspace{0.5cm}

By the notion of composition of functions one may infer that the dynamics of the system $\mc{S}$ may be studied by considering the iteration of the composite function $\Gamma=\phi\circ f$ or $\Phi=f\circ\phi$.  In either case only one single function ($\Gamma$ or $\Phi$) and iteration of one single variable ($x$ or $y$) are involved.  It is indeed true if we focus on the evolution of $x$ or $y$ alone and in this case the dynamical equations \eqref{recur_f} and \eqref{recur_phi} can then be written, in terms of $\Gamma$, as
\be\label{recur_Gamma} x_{i+1}=\Gamma(x_i),\quad i=0,1,2,\dotsc, \ee
or, in terms of $\Phi$, as
\be\label{recur_Phi} y_{i+1}=\Phi(y_i),\quad i=0,1,2,\dotsc .\ee
Note that the orbit of $x$ under $\Gamma$ is the set $\{x,\Gamma(x),\Gamma^{2}(x),\dotsc\}$ and the orbit of $y$ under $\Phi$ is the set $\{y,\Phi(y),\Phi^{2}(y),\dotsc\}$.    However, if we consider only the iteration of $\Gamma$ or $\Phi$, then the interrelationship between $f$ and $\phi$ in determining the system behavior we have just demonstrated, will become opaque.

\defn The real number $\bar{x}$ is called a {\bf fixed point of the composite function} $\Gamma=\phi\circ f$ if
\bes \Gamma(\bar{x})=\phi(f(\bar{x}))=\bar{x}.\ees  Similarly,
$\bar{y}$ is a fixed point of the composite function $\Phi=f\circ \phi$ if
\bes \Phi(\bar{y})=f(\phi(\bar{y}))=\bar{y}.\ees

\vspace{-0.4cm}
\eod

\medskip\noindent It is easy to establish the following fact.

\proposition\label{fixedpoint_rel} If $p=(\bar{x},\bar{y})$ is a fixed point of $\mc{S}$, then $\bar{x}$ is a fixed point of the composite function $\Gamma=\phi\circ f$ and $\bar{y}$ is a fixed point of the composite function $\Phi=f\circ \phi$.

\proof If $p=(\bar{x},\bar{y})$ is a fixed point of $\mc{S}$, then by definition $f(\bar{x})=\bar{y}$ and $\phi(\bar{y})=\bar{x}$.  It follows that $\Gamma(\bar{x})=\phi(f(\bar{x}))=\phi(\bar{y})=\bar{x}$ and $\Phi(\bar{y})=f(\phi(\bar{y}))=f(\bar{x})=\bar{y}$.  \eop

\medskip
In Section \ref{section_model} we obtain from Equations \eqref{recur_f} and \eqref{recur_phi} that the dynamical system $\mc{S}$ will not ``move" if the initial values of $x$ and $y$ coincide with that of a fixed point, i.e., $(x_0,y_0)=(\bar{x},\bar{y})$.  What will happen if it is not the case?   As briefly mentioned also in Section \ref{section_model}, two natural scenarios concerning the {\bf\emph{stability}} of the fixed point emerge.  One is that the state will evolve towards the fixed point and the other is that the state will leave further from the fixed point and may not return.  The trajectories of a tiny ball placed near the bottom of a bowl illustrate very well the first kind of fixed points.  Turning the bowl upside down and trying to keep the ball at the top of the bowl provides a good experience of the second kind of fixed points.  In both cases the curvature of the bowl clearly determines the nature of the ball trajectories and that is why we need differential calculus to study the dynamics of $\mc{S}$.  A precise mathematical description of these two kinds of fixed points is given as follow.

\defn A fixed point $\bar{x}$ of $\Gamma=\phi\circ f$ is called {\bf stable} if, for every $\epsilon >0$, there is a $\delta >0$ such that, for every $x_0$ for which $|x_0-\bar{x}|<\delta$, the $n$-th iteration $\Gamma^{n}(x_0)=\underbrace{\Gamma(\cdots\Gamma(\Gamma}_{n}(x_0)))$ of $x_0$ satisfies $|\Gamma^{n}(x_0)-\bar{x}|<\epsilon$ for every $n\geq 0$.  The fixed point is called {\bf unstable} if it is not stable.  The fixed point is {\bf asymptotically stable} if it is stable and furthermore, there is an $r>0$ such that, for every $x_0$ satisfying $|x_0-\bar{x}|<r$, $\lim_{n\to\infty}\Gamma^{n}(x_0)=\bar{x}$.  An asymptotically stable fixed point is also called an {\bf attracting} fixed point and an unstable fixed point is also called a {\bf repelling} fixed point.\eod

\medskip \noindent The stability of a fixed point $\bar{y}$ of the function $\Phi=f\circ\phi$ is defined exactly in the same way.  The stability of a fixed point of the dynamical system $\mc{S}$ is defined as follow, taking note from Proposition~\ref{fixedpoint_rel} that if $p=(\bar{x},\bar{y})$ is a fixed point of $\mc{S}$, then $\bar{x}$ is a fixed point of $\Gamma$ and $\bar{y}$ is a fixed point of $\Phi$, and vice versa.

\defn\label{def_AS_US_S} A fixed point $p=(\bar{x},\bar{y})$ of $\mc{S}$ is called {\bf stable} ({\bf asymptotically stable}) if both $\bar{x}$ and $\bar{y}$ are stable (asymptotically stable).  The fixed point is called {\bf unstable} if it is not stable.\eod

\medskip
Suppose $f(x)$ and $\phi(y)$ are differentiable functions.  Then their derivatives $f'(x)$ and $\phi'(y)$ exist at every point in their respective domains.  The following result is of vital importance because it tells us how the stability of a fixed point is determined by the cognitive function and the manipulative function.  We will also use this result explicitly in Subsection \ref{subsection_behavior} to study the behavior of a reflexive system.  The practical significance of this result, when the exchange rate example in Section \ref{section_model} is used, is that the discrepancy in the relationship between exchange rate and interest rate near a stable fixed point will remain small and can even be eliminated if the fixed point is asymptotically stable.  On the other hand, the discrepancy can never be narrowed near an unstable fixed point.

\theorem\label{theorem_stab_cond} If $p=(\bar{x},\bar{y})$ is a fixed point of $\mc{S}$ and $f(x)$ and $\phi(y)$ are differentiable functions, then $p$ is asymptotically stable if
\be\label{AS_condition} |f'(\bar{x})|<\frac{1}{|\phi'(\bar{y})|}~,\ee
and it is unstable if
\be\label{US_condition} |f'(\bar{x})|>\frac{1}{|\phi'(\bar{y})|}~.\ee

\proof Since $f(x)$ and $\phi(y)$ are differentiable, then the composite function $\Gamma=\phi\circ f$ is also differentiable (see~\citep{gas&nar98} or other standard texts on calculus).  Furthermore,
\be\label{derivative_comp_fn_G} \Gamma'(x)=\phi'(f(x))f'(x).
\ee
A fixed point $\bar{x}$ of a differentiable function, say our function $\Gamma(x)$, is asymptotically stable if $|\Gamma'(\bar{x})|<1$ and is unstable if $|\Gamma'(\bar{x})|>1$. (This is a central result of discrete dynamical systems; see, for example,~\citep{hal&koc96} or~\citep{hol96}.)  It thus follows from \eqref{derivative_comp_fn_G} and $f(\bar{x})=\bar{y}$ (since $p$ is a fixed point of $\mc{S}$) that $\bar{x}$ is an asymptotically stable fixed point of $\Gamma$ if
\bes
|\phi'(\bar{y})f'(\bar{x})|<1,
\ees
i.e., $|f'(\bar{x})|<\frac{1}{|\phi'(\bar{y})|}$;
and it is unstable if
\bes
|\phi'(\bar{y})f'(\bar{x})|>1,
\ees
i.e., $|f'(\bar{x})|>\frac{1}{|\phi'(\bar{y})|}$.

Now consider the composite function $\Phi=f\circ\phi$, which is differentiable with derivative
\be\label{derivative_comp_fn_P} \Phi'(y)=f'(\phi(y))\phi'(y).
\ee
As for $\Gamma(x)$, a fixed point $\bar{y}$ of $\Phi(y)$ is asymptotically stable if $|\Phi'(\bar{y})|<1$ and is unstable if $|\Phi'(\bar{y})|>1$.  It thus follows from \eqref{derivative_comp_fn_P} and $\phi(\bar{y})=\bar{x}$ (since $p$ is a fixed point of $\mc{S}$) that $\bar{y}$ is an asymptotically stable fixed point of $\Phi$ if
\bes
|f'(\bar{x})\phi'(\bar{y})|<1,
\ees
i.e., $|f'(\bar{x})|<\frac{1}{|\phi'(\bar{y})|}$;
and it is unstable if
\bes
|f'(\bar{x})\phi'(\bar{y})|>1,
\ees
i.e., $|f'(\bar{x})|>\frac{1}{|\phi'(\bar{y})|}$.
We therefore obtain stability conditions \eqref{AS_condition} and \eqref{US_condition} for $\bar{y}$, which are identical to that for $\bar{x}$ when $p=(\bar{x},\bar{y})$ is a fixed point of $\mc{S}$.  In view of Definition \ref{def_AS_US_S} these are also stability conditions for the fixed point $p$ of $\mc{S}$.  The theorem is thus proved.
\eop

\medskip
\subsection{Behavior of A Reflexive System}\label{subsection_behavior}
Recall our discussion in Section \ref{section_model}, that if $f$ and $\phi$ are inverses of each other, then every $x$ in the domain of $f$ is a fixed point of $\Gamma$ and every $y$ in the domain of $\phi$ is a fixed point of $\Phi$.   This is because
\bes \phi(f(x))=f^{-1}(f(x))=x \ees and
\bes f(\phi(y))=\phi^{-1}(\phi(y))=y.\ees
Consequently, every pair $(x,y)$ is a fixed point of the dynamical system $\mc{S}$.  For the exchange rate example we constantly refer to, the existence of this inverse relationship implies that both parties who determine the respective cognitive and manipulative functions agree on certain one-one correspondence between {\bf\emph{all}} interest rates and exchange rates.  In this case a specific exchange rate, once determined for whatever reason, will remain fixed.  However, the theory of reflexivity rejects this extreme scenario and in reality the exchange rate must change with time, not to mention that a function does not necessarily has an inverse.  Indeed, a continuous function is invertible if and only if it is either strictly increasing or decreasing. Therefore the nonexistence of this inverse relationship is a proper assumption, resulting in none or finite number (rather than infinite number) of fixed points that govern the dynamical behavior of the reflexive system concerned.  Figure~\ref{fig_not_inverse} shows an example with three fixed points.  Note that the actual dynamics of $\mc{S}$ can be easily computed via \eqref{recur_f} and \eqref{recur_phi} once $f$ and $\phi$ are given.  In other words, the reflexive system can be studied by computer simulation for various cognitive and manipulative functions without difficulty.  Nevertheless, we would like to take up the following two cases to demonstrate the possible distinct dynamical behaviors when the exact inverse relationship is broken.  This helps us to understand the emergence of Soros's ``boom then bust" phenomenon.

\subsection*{Case 1.~The inverse relationship is mildly violated}  
\begin{center}
\begin{figure}[!h]\centering
\psfrag{x}[][]{$x$}\psfrag{y}[][]{$y$}
\psfrag{x1}[][]{$\bar{x}_1$}\psfrag{y1}[][]{$\bar{y}_1$}
\psfrag{x2}[][]{$\bar{x}_2$}\psfrag{y2}[][]{$\bar{y}_2$}
\psfrag{x3}[][]{$\bar{x}_3$}\psfrag{y3}[][]{$\bar{y}_3$}
\psfrag{f(x)}[][]{$f(x)$}\psfrag{p(y)}[][]{$\phi(y)\neq f^{-1}(y)$}
\psfrag{x0y0}[][]{$(x_0,y_0)$}
\includegraphics[scale=1]{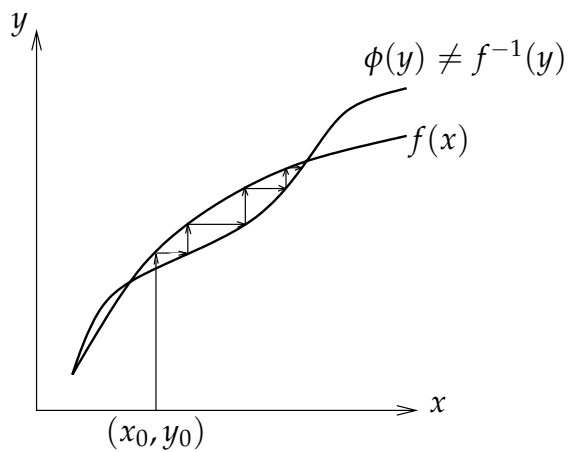}
\caption{Case 1:~$\phi(y)$ is ``approximately equal" to $f^{-1}(y)$.}\label{fig_not_inverse}
\end{figure}
\end{center}

\begin{center}
\begin{figure}[!h]\centering
\psfrag{x}[][]{$x$}\psfrag{y}[][]{$y$}
\psfrag{f(x)}[][]{$f(x)$}\psfrag{p(y)}[][]{$\phi(y)\neq f^{-1}(y)$}
\psfrag{x0y0}[][]{$(x_0,y_0)$}
\includegraphics[scale=1]{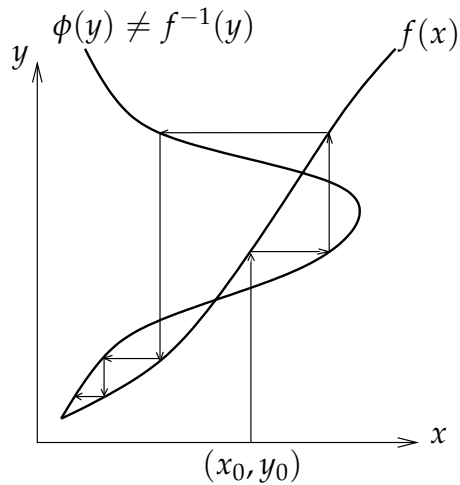}
\caption{Case 2:~$\phi(y)$ is vastly different from $f^{-1}(y)$.}\label{fig_strong_not_inverse}
\end{figure}
\end{center}

In this case $\phi(y)$ is ``approximately equal" to $f^{-1}(y)$.  Note that we here fix $f$ and let $\phi$ to vary starting from $f^{-1}$.  We can also do the opposite and it will not change our results.  To be precise we define a {\bf distance} $d$ (a nonnegative real number) between the two functions and demand that $d$ is not greater than a certain small number.  For example, we can define the distance as
\be\label{distance}
d=\max_{y}|\phi(y)-f^{-1}(y)|,
\ee
i.e., we find the absolute value of the difference between $\phi(y)$ and $f^{-1}(y)$, which is a function having only nonnegative values, then take the maximum value of this function over all possible values of $y$ as the distance.   In this regard, the distance between $\phi(y)$ and $f^{-1}(y)$ in Figure~\ref{fig_not_inverse} may be considered relatively small compared with the two functions in Figure~\ref{fig_strong_not_inverse}.

\vspace{0.5cm}

\vspace{0.5cm}

\begin{center}
\begin{figure}[!h]\centering
\psfrag{x}[][]{$x$}\psfrag{y}[][]{$y$}
\psfrag{p0}[][]{$(x_0,y_0)$}\psfrag{pf}[][]{$(\bar{x},\bar{y})$}
\includegraphics[scale=1]{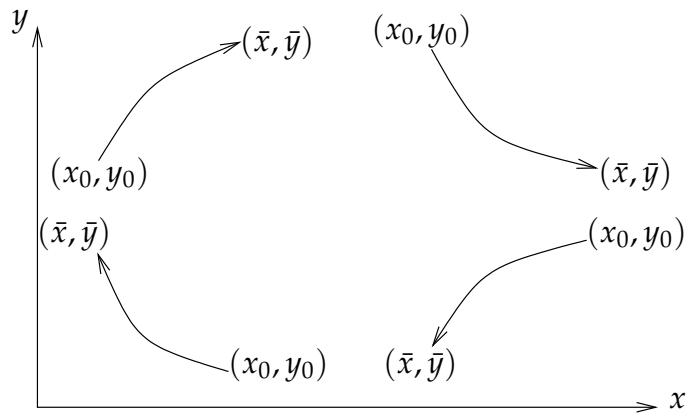}
\caption{The phase portrait for Case 1.}\label{fig_phase_case_1}
\end{figure}
\end{center}

What will be the typical dynamics of $\mc{S}$ when $d$ is small?  Intuitively they should be fairly closed to that when $\phi(y)=f^{-1}(y)$, i.e., disregarding the value of the initial state, the system will converge ``rapidly" towards an asymptotically stable fixed point.  This is shown graphically in Figure~\ref{fig_not_inverse} by using the so-called ``staircase diagram" for plotting the state evolution according to Equations \eqref{recur_f} and \eqref{recur_phi}.  The diagram is a commonly used graphical method to obtain the dynamics of a recursive system.  See~\citep{hal&koc96} or~\citep{hol96} for details.  On the other hand we can also depict this evolution of system state with time on the $x$-$y$ plane.  The plot is commonly called the {\bf phase portrait} of the system.  Several examples of this portrait for the dynamics we have just referred to are shown in Figure~\ref{fig_phase_case_1}.\endnote{Though the portraits should be plotted as sequences of points since $\mc{S}$ is a discrete-time system, we use solid lines passing these points for easy visualization of the direction of state movements.}  Observe that in all these examples the system state evolves from its initial value $(x_0,y_0)$ towards a fixed point $(\bar{x},\bar{y})$ {\bf\emph{monotonically}}.  A non-rigorous explanation of this pattern of dynamics is as follow.

\vspace{0.5cm}

Suppose $\phi(y)$ is a ``small" variation of $f^{-1}(y)$, then $\mc{S}$ is likely to have fixed points of alternating stability signatures along the $x$ or $y$ axis, i.e., one attracting fixed point followed by a repelling fixed point, then another attracting fixed point, and so on.  This is because the relationship between the derivatives of $\phi$ and $f$ are likely to switch from satisfying one condition, for example \eqref{AS_condition}, to another condition \eqref{US_condition} and then back to \eqref{AS_condition} along the fixed points.  The result is an alternation of fixed-point types according to Theorem~\ref{theorem_stab_cond}.  Figure~\ref{fig_not_inverse} provides a typical example and we can easily identify this alternation by examining the derivatives of $\phi$ and $f$ at the three consecutive fixed points.  Then $\mc{S}$ will exhibit very simple dynamics like that shown in Figure~\ref{fig_not_inverse} in which an initial state will be repelled by an unstable fixed point and at the same time attracted towards a nearby stable fixed point.
\subsection*{Case 2.~The inverse relationship is strongly violated}  
\begin{center}
\begin{figure}[!h]\centering
\psfrag{x}[][]{$x$}\psfrag{y}[][]{$y$}
\psfrag{p0}[][]{$(x_0,y_0)$}\psfrag{pf}[][]{}
\includegraphics[scale=1]{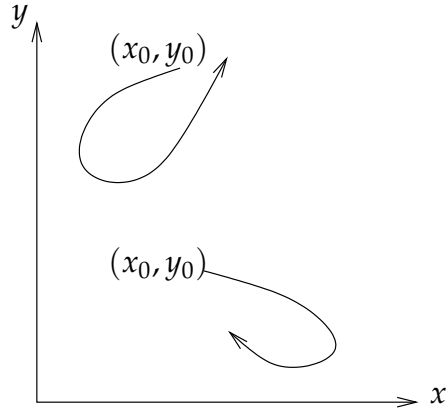}
\caption{The phase portrait for Case 2.}\label{fig_phase_case_2}
\end{figure}
\end{center}

In this case $\phi(y)$ and $f^{-1}(y)$ are vastly different and the distance $d$ between them is of large value.  In real terms the parties who elect their respective cognitive and manipulative functions disagree strongly at least in certain range of market values such as that shown in the upper part of Figure~\ref{fig_strong_not_inverse}.  This can lead to, as demonstrated by Birshtein and Borsevici in~\citep{bir&bor02}, the so-called ``boom then bust" phenomenon Soros frequently refers to.  The key characteristic of this phenomenon is an obvious reversal of price at some point such as that depicted in the staircase diagram in Figure \ref{fig_strong_not_inverse} or the phase portrait Figure~\ref{fig_phase_case_2}.  The {\bf\emph{return}} of price back to its starting value or the neighborhood of this starting value can be modeled by the following notions in dynamical systems theory.

\vspace{0.5cm}

\defn Given a function $f$ a point $p$ is said to be {\bf recurrent} for $f$ if, for any open interval $J$ containing $p$, there exists $n>1$ such that $f^{n}(p)\in J$.  When $f^{n}(p)$ returns exactly to $p$ we call $p$ a {\bf periodic point} with {\bf period} $n$.\eod

\rems In the case of our reflexive system $\mc{S}$, a recurrent point or a periodic point is a pair of real numbers and the open interval becomes an open set on the $x$-$y$ plane.  It is in general impossible to calculate the periodic points if any.  Practical functions may even preclude periodic points completely since they are far from hypothetical functions which generate periodic points.  On the other hand, the condition for the existence of recurrent points is deemed to be less stringent.

\bigskip
The marked difference between the two behaviors of a reflexive system as described in the above Cases 1 and 2 can be made precise by using the following concepts in dynamical systems theory~(\citep{dev89},\citep{hol96},\citep{rob04}).
\defn A function is called a {\bf homeomorphism} if it is continuous, has an inverse, and furthermore, the inverse is also continuous.
\eod
\defn\label{topo_conj_defn} Let $f\colon A\rightarrow A$ and $g\colon B\rightarrow B$ be two functions.  Then $f$ is {\bf topologically conjugate} to $g$ if there exists a homeomorphism $h$ from $A$ to $B$ such that $h\circ f=g\circ h$.  In this case $h$ is called a {\bf topological conjugacy}. \eod

\medskip A topological conjugacy can be regarded as a transformation of coordinates or transformation of functions.  Its importance can be easily seen from the following considerations.  First, we note that, for every $x$ in the domain of $f$,
\bes
h(f(x))=g(h(x)).
\ees
It follows that if $\bar{x}$ is a fixed point of $f$, then
\bes h(f(\bar{x}))=h(\bar{x})=g(h(\bar{x})),
\ees
which means $h(\bar{x})$ is a fixed point of $g$.  Indeed, $f$ and $g$ induce {\bf\emph{identical}} recursive dynamics if they are topological conjugate, in the sense that orbits under $f$ and $g$ are related to each other simply by a homeomorphism, namely,
\bes
y_i=h(x_i),\quad i=0,1,2,\dotsc .
\ees
This important result (its proof can be found in \citep{rob04}) leads to a further concept called the {\bf structural stability} of a dynamical system by considering $g$ as a perturbation (or an approximation) of $f$.  The central question to ask is whether $f$ and $g$ are topological conjugate if the distance between the two functions (such as that defined in \eqref{distance}) is small.

Thus, using the language of topological conjugacy, we venture that the reflexive system $\mc{S}_1$ in Case 1 is {\bf\emph{not}} topologically conjugate to the reflexive system $\mc{S}_2$ in Case 2.  Here the homeomorphism, which cannot be found to effect conjugacy, is a function from $\R^2$ to $\R^2$.

\section{Conclusions}
It might be Soros's intention to formalize his theory of reflexivity when he introduced the recursive equation~\eqref{reflex_eqns} and its governing functions $f$ and $\phi$.  Nevertheless, without a formal mathematical analysis of the equation, it remains indeterminate whether the equation does provide an effective modeling of reflexivity for the purpose of explanation at the minimum and prediction at best.  Astonishingly there has been no serious effort to conduct this relatively easy (from the mathematical point of view) analysis to answer this basic question.  We do not claim we have given a complete answer here.  Rather, we believe that we have elucidated, through the use of a simple exchange rate example and ``naive" mathematics: (i) The essence of Soros's theory of reflexivity as modeled by this recursive equation; (ii) The importance of the notion of fixed points pertaining to the dynamical behavior of the equation; (iii) The idea of stability of a fixed point and how it is deduced from the given cognitive and manipulative functions; and (iv) The emergence of equilibria and ``boom then bust" as qualitative behaviors of this equation.

However, a fundamental issue stays unsettled, namely, by what means can we determine the two governing functions $f$ and $\phi$, even we believe that the theory of reflexivity can be applied to our problem?  Soros himself has given no explicit form of $f$ and $\phi$ though he has used, during the past twenty years, many examples extracted from finance, history, and political science to explain the possible dynamics generated by the two functions, in particular phenomena exhibiting ``boom then bust".  We have drawn an example proposed by Birshtein and Borsevici~\citep{bir&bor02}, that this kind of behavior can indeed be generated by some simple pair of cognitive and manipulative functions.  Whether the functions could be inferred from historical data or derived from ``expert opinions" is subject to further exploration and constitutes a worthwhile research direction.  Would our exchange rate example offer an experimental platform for such a possibility?  However, the problem will become acute when the reflexive system contains more than two variables---an extension suggested by Soros for his theory of reflexivity~\citep{sor00}.  In this case the cognitive and the manipulative functions are functions of multi variables.

\newpage
\theendnotes

\end{document}